\documentclass[3p,twocolumn]{elsarticle}

\usepackage{amsmath}  
\usepackage{amsfonts}
\usepackage{amssymb}
\usepackage{natbib} 
\usepackage{setspace}
\usepackage{graphicx} 
\usepackage{xspace}
\usepackage{cancel}
\usepackage{url}
\usepackage{color}

\newcommand{\mi}{\mathrm{i}} 

\newcommand{\dfd}[3]{\hspace{-0.4em}\ensuremath{\frac{\mathrm{d}^{#1}#3}{(2\pi)^{#2}}}\,}

\newcommand{\eqn}[1]{Eq.~(\ref{#1})}

\newcommand{\tab}[1]{Tab.~\ref{#1}}
\newcommand{\sect}[1]{Sec.~\ref{#1}}

\def\Kbar{\overline{K}}
\def\K0bar{\overline{K^0}}

\def\qbar{\bar{q}}

\def\Kbar{\overline{K}}

\newcommand{\be}{\begin{equation}}
\newcommand{\ee}{\end{equation}}
\newcommand{\bge}{\begin{equation}}
\newcommand{\ene}{\end{equation}}
\newcommand{\bea}{\begin{eqnarray}}
\newcommand{\eea}{\end{eqnarray}}
\newcommand{\bg}{\begin{eqnarray}}
\newcommand{\en}{\end{eqnarray}}

\begin{document}





\title{
  \vspace{-30mm}
  \begin{flushright} ADP-17-11/T1017 \end{flushright}
  \vspace{23mm}
$\phi$ meson mass and decay width in nuclear matter and nuclei}

\author[add1,add4]{J.J.~Cobos-Mart\'{\i}nez\corref{corr}}
\ead{javiercobos@ifm.umich.mx}
\author[add1]{K.~Tsushima}
\ead{kazuo.tsushima@gmail.com}
\author[add2]{G.~Krein}
\ead{gkrein@ift.unesp.br}
\author[add3]{A.W.~Thomas}
\ead{anthony.thomas@adelaide.edu.au}

\address[add1]{Laborat\'orio de F\'{\i}sica Te\'orica e Computacional - LFTC, 
Universidade Cruzeiro do Sul, 01506-000, S\~ao Paulo, SP, Brazil}
\address[add2]{Instituto de F\'{\i}sica Te\'orica, Universidade Estadual 
Paulista, Rua Dr. Bento Teobaldo Ferraz, 271 - Bloco II, 01140-070, 
S\~ao Paulo, SP, Brazil}
\address[add3]{CSSM and ARC Centre of Excellence for Particle 
Physics at the Terascale, Department of Physics, University of Adelaide,
Adelaide SA 5005, Australia}
\address[add4]{Instituto de F\'{\i}sica y Matem\'aticas, Universidad Michoacana 
de San Nicol\'as de Hidalgo, Edificio C-3, Ciudad Universitaria, Morelia, 
Michoac\'an 58040, M\'exico}

\cortext[corr]{Corresponding author}

\begin{abstract}
The mass and decay width of the $\phi$ meson in cold 
nuclear matter are computed 
in an effective Lagrangian approach. The medium dependence of these 
properties are obtained by evaluating kaon-antikaon loop contributions to the 
$\phi$ self-energy, employing the medium-modified kaon masses, 
calculated using the quark-meson coupling model. The loop integral 
is regularized with a dipole form factor, and the sensitivity of 
the results to the choice of  
cutoff mass in the form factor is investigated. At normal nuclear matter density 
we find a downward shift of the $\phi$  mass by a few percent, while the 
decay width is enhanced by an order of magnitude. For a large variation of the 
cutoff mass parameter, the results for the $\phi$ mass and the decay width 
turn out to vary very little. Our results support results in 
the literature which suggest that one should observe 
a small downward mass shift and a large broadening of the decay width. 
In order to explore the possibility of studying the binding and 
absorption of $\phi$ mesons  
in nuclei, we also present the single-particle 
binding energies and half-widths of $\phi$-nucleus bound states 
for some selected nuclei. 
\end{abstract}


\maketitle

\section{Introduction}

The study of the changes in light vector meson properties in a nuclear medium 
have attracted much experimental and theoretical interest 
{\textemdash}see Refs.~\cite{Leupold:2009kz,Hayano:2008vn,{Krein:2016fqh}} 
for recent reviews. 
Amongst the arguments motivating these studies we mention  
the interest in chiral 
symmetry restoration at high density and the possible role of QCD van der Waals 
forces.
In particular, there is special interest on the $\phi$ meson, 
the main reasons being: 
(i)~despite its nearly pure~$s\overline{s}$ content, the $\phi$ does
interact strongly with a nucleus, composed predominantly of  
light $u$ and $d$ quarks, through 
the excitation of below-threshold virtual kaon and anti-kaon states 
that might have their properties 
changed in medium, the latter issue in itself being also of current 
interest~\cite{Tsushima:1997df,Laue:1999yv,SchaffnerBielich:1999cp,Akaishi:2002bg,Fuchs:2005zg};  
(ii) the  $\phi N$ interaction in 
vacuum~\cite{Titov:1997qz,Titov:1998bw,Oh:1999nv,Oh:2001bq} 
and a possible in-medium mass shift of the $\phi$ are related to 
the strangeness content of the 
nucleon~\cite{Gubler:2014pta}, which may have implications beyond 
the physics of the strong interaction, 
affecting, for example, the experimental searches for dark 
matter~\cite{Bottino:2001dj,Ellis:2008hf,Giedt:2009mr}; 
(iii) medium modifications of $\phi$ properties have been 
proposed~\cite{Sibirtsev:2006yk}
as a possible source for the anomalous nuclear mass number $A$-dependence observed 
in $\phi$ production from nuclear 
targets~\cite{Ishikawa:2004id}; (iv) furthermore, as the $\phi$ is a nearly 
pure $s\overline{s}$ state and gluonic interactions are flavor blind,
studying it serves to test theories of the 
multi-gluon exchange interactions, including long range QCD van der 
Waals forces~\cite{Appelquist:1978rt}, which are believed to play a 
role in the binding 
of the J/$\Psi$ and other exotic heavy-quarkonia to 
matter~\cite{Brodsky:1989jd,Luke:1992tm,Sibirtsev:1999jr,Gao:2000az,Beane:2014sda,Brambilla:2015rqa,Gao:2017hya,Kawama:2014pja,Kawama:2014iwa,Ohnishi:2014xla,
Aoki:2015qla,Morino:2015bqa}.

Heavy-ion collisions and photon- or proton-induced reactions on nuclear 
targets have been used to extract information on the in-medium properties 
of hadrons. Although the medium modifications of hadron properties 
are expected to be stronger in heavy-ion collisions, they are also expected 
to be large enough in photon- or proton-induced reactions to enable the study  
of in-medium properties by fixed-target experiments. 
Several experiments have focused 
on the light vector mesons $\rho$, $\omega$, 
and $\phi$, since their mean-free paths can be comparable with the size of 
a nucleus after being produced inside the nucleus. 
However, a unified consensus has not yet been reached among the different 
experiments{\textemdash}see Refs.~\cite{Leupold:2009kz,Hayano:2008vn,{Krein:2016fqh}} for 
comprehensive reviews of the current status. 

For the $\phi$ meson, although the precise values are different, 
a large in-medium broadening of the width has been reported 
by most of the experiments 
performed, while only a few of them find evidence 
for a substantial mass shift.
For example, the KEK-E325 collaboration~\cite{Muto:2005za} reported a mass 
reduction of $3.4\%$ and an in-medium decay width of $\approx 14.5$ MeV at 
normal nuclear matter density.
The latter disagrees with the SPring8~\cite{Ishikawa:2004id} result, which 
reported a large in-medium $\phi N$ cross section leading to a decay
width of 35 MeV. But this 35 MeV is in close agreement with the two JLab CLAS  
collaboration measurements reported 
in Refs.~\cite{Mibe:2007aa} and~\cite{Qian:2009ab}. 

In an attempt to clarify the situation, 
the CLAS collaboration at JLab~\cite{Wood:2010ei} 
performed new measurements of nuclear transparency ratios, 
and estimated in-medium widths in the range of 23-100 MeV. 
These values overlap with that of the SPring8 measurement~\cite{Ishikawa:2004id}.
More recently, the ANKE-COSY collaboration~\cite{Polyanskiy:2010tj} has 
measured the $\phi$ meson production from proton-induced reactions on various 
nuclear targets. 
The comparison of data with model calculations suggests an in-medium 
$\phi$ width of $\approx 50$ MeV. 
This result is consistent with that of SPring8~\cite{Ishikawa:2004id},  
as well as the one deduced from CLAS at JLab~\cite{Wood:2010ei}. 
However, the value is clearly larger than that of the KEK-E325 
collaboration~\cite{Muto:2005za}.

From the discussions above, it is obvious that the search for evidence of 
a light vector meson mass shift is indeed complicated. 
It certainly requires further experimental 
efforts to understand better the changes of $\phi$ properties  
in a nuclear medium. 
For example, the J-PARC E16 collaboration~\cite{JPARCE16Proposal} intends to 
perform a more systematic study for the mass shift of vector mesons 
with higher statistics.
Furthermore, the E29 collaboration at J-PARC has recently put forward a 
proposal~\cite{JPARCE29Proposal,JPARCE29ProposalAdd} to study the 
in-medium mass modification of $\phi$ via the possible formation of 
the $\phi$-nucleus bound states~\cite{Aoki:2015qla}, 
using the primary reaction $\overline{p}p\rightarrow \phi\phi$. 
Finally, there is a proposal at JLab, following the 12 GeV upgrade, to study 
the binding of $\phi$ (and $\eta$) to $^4$He~\cite{JLabphi}.

On the theoretical side, various authors predict a downward shift of the 
in-medium $\phi$ meson mass and a broadening of the decay width.
The possible decrease of the light vector meson masses in a nuclear medium 
was first predicted by Brown and Rho~\cite{Brown:1991kk}.
Thereafter, many theoretical investigations have been conducted, some of 
them focused on the self-energies of the $\phi$ due to the kaon-antikaon loop.
Ko et al.~\cite{Ko:1992tp} used a density-dependent kaon mass determined from
chiral perturbation theory and found that at normal nuclear 
matter density, $\rho_0$,
the $\phi$ mass decreases very little, by at most $2\%$, and the width 
$\Gamma_\phi \approx 25$~MeV and broadens drastically for large densities. 
Hatsuda and Lee calculated the in-medium 
$\phi$ mass based on QCD sum rule approach~\cite{Hatsuda:1991ez,Hatsuda:1996xt}, 
and predicted a decrease of 1.5\%-3\% at normal nuclear matter density.
Other investigations also predict a large broadening 
of the $\phi$ width: Ref.~\cite{Klingl:1997tm} reports a 
negative mass shift of $ < 1\%$ and a decay width of 45 MeV at $\rho_0$;   
Ref.~\cite{Oset:2000eg} predicts a decay width of 22 MeV but does not report 
a result on the mass shift; and Ref.~\cite{Cabrera:2002hc} gives a rather 
small negative mass shift of $\approx 0.81\%$ and a decay width of 30 MeV.
More recently, Ref.~\cite{Gubler:2015yna} reported a downward mass shift of
$< 2\%$ and a large broadening width of 45 MeV; 
and finally, in Ref.~\cite{Cabrera:2016rnc}, 
extending the work of Refs.~\cite{Oset:2000eg,Cabrera:2002hc}, the authors reported 
a negative mass shift of $3.4\%$ and a large decay width of 70 MeV at $\rho_0$.
The reason for these differences may lie in the different approaches 
used to estimate the kaon-antikaon loop contributions for the $\phi$ self-energy.

In the present article we report results for the $\phi$ mass shift
and decay width in nuclear matter, taking into account 
the medium dependence of the $K$ and 
$\Kbar$ masses. The latter are included by an explicit calculation 
based upon the quark-meson 
coupling (QMC) model~\cite{Guichon:1989tx,Guichon:1995ue}. 
The QMC model is a quark-based 
model of finite nuclei and nuclear matter, and has been very successful 
in describing the nuclear matter saturation properties, hadron properties in 
nuclear medium, as well as the properties of finite 
nuclei~\cite{Stone:2016qmi} and hypernuclei~\cite{qmchyp}{\textemdash}for 
a comprehensive review of the QMC model, see Ref.~\cite{Saito:2005rv}.

The paper is organized as follows. In~\sect{sec:vacse} we present the effective
Lagrangian used to calculate the $\phi$-meson self-energy in vacuum, and give 
explicit expressions for its real and imaginary parts. Since the in-medium properties 
of the $\phi$ are dependent on the kaon and anti-kaon masses in a nuclear medium  
calculated within the QMC model, we briefly review this 
model in~\sect{sec:qmcmodel}, and 
provide the necessary detail to understand the dressing of the kaons in nuclear 
medium. In~\sect{sec:medse} we calculate the $\phi$-meson 
self-energy in nuclear matter    
and report the in-medium $\phi$-meson mass and decay width, as well as the binding 
energies and widths of selected $\phi$-nucleus bound states. 
Finally, conclusions and perspectives are given 
in~\sect{sec:conclusions}.

\section{\label{sec:vacse} $\phi$ meson self-energy in vacuum}
We use the effective Lagrangian of Refs.~\cite{{Ko:1992tp},{Klingl:1996by}}
to compute the $\phi$ self-energy; 
the interaction Lagrangian $\mathcal{L}_{int}$ involves 
$\phi K\Kbar$ and $\phi\phi K\Kbar$ couplings dictated by a 
local gauge symmetry principle:
\begin{equation}
 \label{eqn:Lint}
\mathcal{L}_{int} = \mathcal{L}_{\phi K\Kbar} + \mathcal{L}_{\phi\phi K\Kbar},
\end{equation}
where
\begin{equation}
\label{eqn:phikk}
\mathcal{L}_{\phi K\Kbar} = \mi g_{\phi}\phi^{\mu}
\left[\Kbar(\partial_{\mu}K)-(\partial_{\mu}\Kbar)K\right],
\end{equation}
and
\begin{equation}
\label{eqn:phi2kk}
\mathcal{L}_{\phi\phi K\Kbar} = g^2_{\phi} \phi^\mu\phi_\mu \Kbar K.
\end{equation}
We use the convention:
\begin{equation}
\label{eqn:isospin}
K=\left(\begin{array}{c} K^{+} \\ K^{0} \end{array} \right),\;
\overline{K}=\left(K^{-}\;\overline{K}^{0}\;\right).
\end{equation}
%
\begin{figure}[t]
\centering
\includegraphics[scale=0.75]{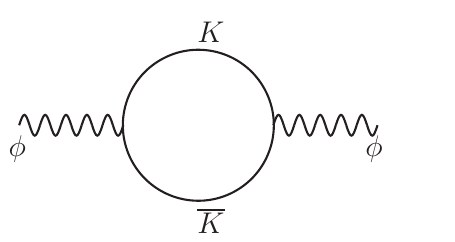}
\caption{\label{fig:phise} $K\Kbar$-loop contribution to the $\phi$ meson 
self-energy.}
\end{figure}
  
We note that the use of the effective interaction Lagrangian of \eqn{eqn:Lint} without the term
given in \eqn{eqn:phi2kk} may be considered as being motivated by the hidden gauge approach
in which there are no four-point vertices, such as \eqn{eqn:phi2kk}, that involve two
pseudoscalar mesons and two vector mesons~\cite{Lin:1999ve,Lee:1994wx}.
This is in contrast to the approach of using the minimal substitution to
introduce vector mesons as gauge particles where such four-point vertices do appear.
However, these two methods have been shown to be consistent if both the vector and
axial vector mesons are included~\cite{Yamawaki:1986zz,Meissner:1986tc,Meissner:1987ge,
Saito:1987ba}.
Therefore, we present results with and without such an interaction.  
We consider first the contribution from the 
$\phi K\Kbar$ coupling given by Eq.~(\ref{eqn:phikk}) to 
the scalar part of the $\phi$ self-energy, 
$\Pi_{\phi}(p)$; Figure~\ref{fig:phise} depicts this contribution. 
{}For a $\phi$ meson at rest, it is given
by
\begin{equation}
\label{eqn:phise}
\mi\Pi_{\phi}(p)=-\frac{8}{3}g_{\phi}^{2}\int\dfd{4}{4}{q}\vec{q}^{\,2}
D_{K}(q)D_{K}(q-p) \, ,
\end{equation}
where $D_{K}(q)=\left(q^{2}-m_{K}^{2}+\mi\epsilon\right)^{-1}$ is the
kaon propagator;  $p=(p^{0}=m_{\phi},\vec{0})$ is 
the $\phi$ meson four-momentum vector, 
with $m_{\phi}$ the $\phi$ meson mass; $m_{K} (=m_{\Kbar})$ is the kaon mass. 
When $m_{\phi}<2m_{K}$ the self-energy $\Pi_{\phi}(p)$ is real. However, when 
$m_{\phi}>2m_{K}$, which is the case here, $\Pi_{\phi}(p)$ acquires an imaginary
part. The mass of the $\phi$ is determined from the real part of $\Pi_{\phi}(p)$
\begin{equation}
\label{eqn:phimassvacuum}
m_{\phi}^{2}=\left(m_{\phi}^{0}\right)^{2}+\Re\Pi_{\phi}(m_{\phi}^{2}),
\end{equation}
with $m_{\phi}^{0}$ being the bare mass of the $\phi$ and
\begin{equation}
\label{eqn:repiphi}
\Re\Pi_{\phi}=-\frac{2}{3}g_{\phi}^{2} \, \mathcal{P}\!\!
\int\dfd{3}{3}{q}\vec{q}^{\,2}\frac{1}{E_{K}(E_{K}^{2}-m_{\phi}^{2}/4)} \, .
\end{equation}
Here $\mathcal{P}$ denotes the Principal Value part of the 
integral Eq.~(\ref{eqn:phise}) 
and $E_{K}=(\vec{q}^{\,2}+m_{K}^{2})^{1/2}$.  
The decay width of $\phi$ to a $K\Kbar$ pair 
is given in terms of the imaginary part of $\Pi_{\phi}(p)$
\begin{equation}
\Im\Pi_{\phi} = - \frac{g_{\phi}^{2}}{24\pi}
m^2_\phi \left(1-\frac{4m_{K}^{2}}{m_{\phi}^{2}}\right)^{3/2},
\end{equation}
as
\begin{equation}
\label{eqn:phidecaywidth}
\Gamma_{\phi} = -\frac{1}{m_{\phi}}\Im\Pi_{\phi} = \frac{g_{\phi}^{2}}{24\pi} m_\phi
\left(1-\frac{4m_{K}^{2}}{m_{\phi}^{2}}\right)^{3/2} \, .
\end{equation}

The integral in \eqn{eqn:repiphi} is divergent and needs 
regularization; we use a phenomenological 
form factor, with a cutoff parameter $\Lambda_{K}$, 
as in Ref.~\cite{Krein:2010vp}. The coupling 
constant $g_{\phi}$ is determined by the experimental 
width of the $\phi$ in vacuum~\cite{PDG:2015}.
{}For the $\phi$ mass, $m_{\phi}$, we use its experimental value: 
$m_{\phi}^{\text{expt}}=1019.461$ MeV~\cite{PDG:2015}. For 
the kaon mass $m_{K}$, there is a small 
ambiguity since $m_{K^+}\ne m_{K^0}$, as a result of charge symmetry breaking and 
electromagnetic interactions. The experimental values for 
the $K^{+}$ and $K^{0}$ meson masses in vacuum are 
$m_{K^{+}}^{\text{expt}}=493.677$ MeV and $m_{K^{0}}^{\text{expt}}=497.611$ MeV, 
respectively~\cite{PDG:2015}. 
For definiteness we use the average of $m_{K^{+}}^{\text{expt}}$ and 
$m_{K^{0}}^{\text{expt}}$ as the value of $m_{K}$ in vacuum.
The effect of this tiny mass ambiguity on the in-medium  
kaon (antikaon) properties is negligible.    
Then, we get the coupling $g_{\phi}=4.539$, and can fix  
the bare mass~$m_{\phi}^{0}$.

\section{\label{sec:qmcmodel}The quark-meson coupling model and the 
in-medium kaon mass}
Essential to our results for the in-medium $\phi$ mass, $m^*_\phi$, and 
decay width, $\Gamma_{\phi}^{*}$, at finite baryon density $\rho_B = \rho_p + \rho_n$ 
(sum of the proton and neutron densities), 
is the in-medium kaon mass, $m_{K}^{*}$, which is driven by the interactions 
of the kaon with the nuclear medium{\textemdash}we denote with an asterisk 
an in-medium quantity. The in-medium kaon mass is calculated in 
the QMC model. This model has been successfully applied 
to investigate the properties of infinite nuclear matter and finite nuclei. 
Here we briefly present the necessary details needed to understand our results.
For a more in depth discussion of the model see Refs.~\cite{Tsushima:1997df,
Guichon:1989tx,Saito:2005rv} and references therein.

We consider nuclear matter in its rest frame, where all the scalar and vector mean 
field potentials, which are responsible for the nuclear many-body interactions,  
are constants in Hartree approximation. 
The Dirac equations for the quarks and antiquarks 
($q = u$ or $d$, and $s$) in a hadron bag in nuclear matter at the position
$x=(t,\vec{r}) \ ({\rm with}~|\vec{r}|\le R^*_h, {\rm the~in~medium~bag~radius})$ 
are given 
by~\cite{Tsushima:1997df,Saito:2005rv}:
\begin{eqnarray}
&&\left[ \mi\cancel{\partial}_x - m^*_q \mp \gamma^0
V_+  \right]
\left( \!\!\begin{array}{c} \psi_u  \\
\psi_{\bar{u}}\end{array} \!\!\right) = 0,
\label{diracu} \nonumber\\[0.25true cm]
&&\left[ \mi\cancel{\partial}_x - m^*_q \mp \gamma^0
V_- \right]
\left( \!\!\begin{array}{c} \psi_d  \\
\psi_{\bar d} \end{array} \!\!\right) = 0,
\label{diracd} \nonumber\\[0.25true cm]
&&\left[ \mi\cancel{\partial}_x - m_{s} \right]
\left( \!\!\begin{array}{c} \psi_s  \\
\psi_{\bar s} \end{array} \!\!\right) = 0, 
\label{diracQ}
\end{eqnarray}
where $m^*_q = m_q - V^q_\sigma$ and $V_{\pm} = V^q_\omega \pm 1/2\,V^q_\rho$. 
Here we neglect the Coulomb force, and assume SU(2) 
symmetry for the light quarks ($m_{q}=m_{u}=m_{d}$). 
The constant mean-field potentials in nuclear matter are defined by 
$V^q_\sigma \equiv g^q_\sigma \sigma$, $V^q_\omega \equiv g^q_\omega \omega$, and 
$V^q_\rho \equiv g^q_\rho b$, where $b$ is the time component of 
the $\rho$ mean field, with $g^q_\sigma$, $g^q_\omega$, and $g^q_\rho$ the 
corresponding quark-meson coupling constants. 
Note that $V^q_\rho \propto (\rho_p - \rho_n) = 0$ in symmetric nuclear matter, 
although this is not true in a nucleus where the Coulomb force may induce  
an asymmetry between the proton and neutron distributions  
even in a nucleus with the same 
number of protons and neutrons, resulting in 
$V^q_\rho \propto (\rho_p - \rho_n) \ne 0$ at a given position in a nucleus.

The normalized, static solution for the ground-state quarks or antiquarks 
with flavor $f$ in the hadron $h$ may be written as 
$\psi_f (x) = N_f e^{- i \epsilon_f t / R_h^*} \psi_f (\vec{r})$,
where $N_f$ and $\psi_f(\vec{r})$ are the normalization factor and 
the corresponding spin and spatial part of the wave function,
respectively. The in-medium bag radius $R_h^*$ of hadron $h$  
is determined through the stability condition for the mass of the 
hadron against the variation of the bag radius~\cite{Guichon:1989tx,
Saito:2005rv}{\textemdash}see Eq.~(\ref{bagRs}) below. 
The eigenenergies in units 
of $1/R_h^*$ are given by
%
\begin{eqnarray}
\left( \!\!\begin{array}{c}
\epsilon_u \\
\epsilon_{\bar{u}}
\end{array} \!\!\right)
&=& \Omega_q^* \pm R_h^* \, V_+,
\\[0.25true cm]
\left( \!\!\begin{array}{c} \epsilon_d \\
\epsilon_{\bar{d}}
\end{array} \!\!\right)
&=& \Omega_q^* \pm R_h^* \, V_-,
\\[0.25true cm]
\epsilon_{s}
&=& \epsilon_{\bar{s}} =
\Omega_{s}.
\label{energy}
\end{eqnarray}
Recall that $V^q_\rho = 0$, as explained earlier. 
The in-medium hadron mass, $m^*_h$, 
is calculated by
\begin{eqnarray}
\hspace{-0.5cm}
&&m_h^* = \sum_{j=q,\bar{q},s,{\bar s}}
\frac{ n_j\Omega_j^* - z_h}{R_h^*}
+ \frac{4\pi}{3} R_h^{* 3} B,
\label{hmass}
\\[0.25true cm]
\hspace{-0.5cm}
&&\frac{\partial m_h^*}
{\partial R^*_h} = 0,
\label{bagRs}
\end{eqnarray}
where $\Omega_q^*=\Omega_{\bar{q}}^* =[x_q^2 + (R_h^* m_q^*)^2]^{1/2}$ 
with $\Omega_s^*=\Omega_{\bar s}^*=
[x_s^2 + (R_h^* m_s)^2]^{1/2}$, $x_{q,s}$ being the lowest bag 
eigenfrequencies; and $n_q (n_{\qbar})$, $n_s (n_{\bar s})$ are the quark 
(antiquark) numbers for the quark flavors $q$ and $s$, respectively. 
The MIT bag quantities, $z_h$, $B$, $x_{q,s}$, and $m_{q,s}$ are the 
parameters for the sum of the c.m.~and gluon fluctuation effects, bag 
constant, lowest eigenvalues for the quarks $q$ or $s$, respectively, 
and the corresponding current quark masses. 
The parameters $z_N$ ($z_h$) and $B$  are fixed by fitting the nucleon 
(hadron) mass in free space. 

For the current quark masses relevant for this study, 
we use $(m_{u,d},m_s) = (5,250)$ MeV, where these values were 
used in Refs.~\cite{Tsushima:1997df,Saito:2005rv} and many studies made in 
the standard version of the QMC model. Since the effects of the current-quark 
mass values on the final results are very small, we use the same values as those 
used in the past, so that we can compare and discuss the results with those 
obtained previously. The bag radius of the nucleon in vacuum 
is taken to be $R_N = 0.8$ fm, and the parameter $z_N$, simulating 
the zero-point and c.m.~energy, is obtained $z_N = 3.295$. For the kaon, 
the values in vacuum calculated here are  
($R_K,z_K$) $=$ (0.574 fm, 3.295).
The bag constant calculated for the present study 
is $B = (170$ MeV$)^4$. The quark-meson 
coupling constants, which are determined so as to reproduce 
the saturation properties of 
symmetric nuclear matter{\textemdash}the binding energy per nucleon of 15.7 MeV at 
$\rho_0 = 0.15$ fm$^{-3}${\textemdash}are 
($g^q_\sigma, g^q_\omega, g^q_\rho$) = ($5.69, 2.72, 9.33$).
In addition, the incompressibility obtained is $K = 279.3$ MeV.
The $\sigma$ coupling at the nucleon level, which is not trivial, is related by 
$g_\sigma \equiv g^N_\sigma \equiv 3 g^q_\sigma S_N(\sigma=0) 
= 3 \times 5.69 \times 0.483 = 8.23$~\cite{Tsushima:1997df,Saito:2005rv}, 
where  
\begin{equation}
S_N(\sigma) = \int d^3r \ \bar{\psi}_q(\vec{r}) \psi_q(\vec{r}),
\end{equation}
with the ground state light-quark wave functions evaluated self-consistently 
in-medium.

%
%
%

\begin{figure}[ht]
\centering
\includegraphics[scale=0.3]{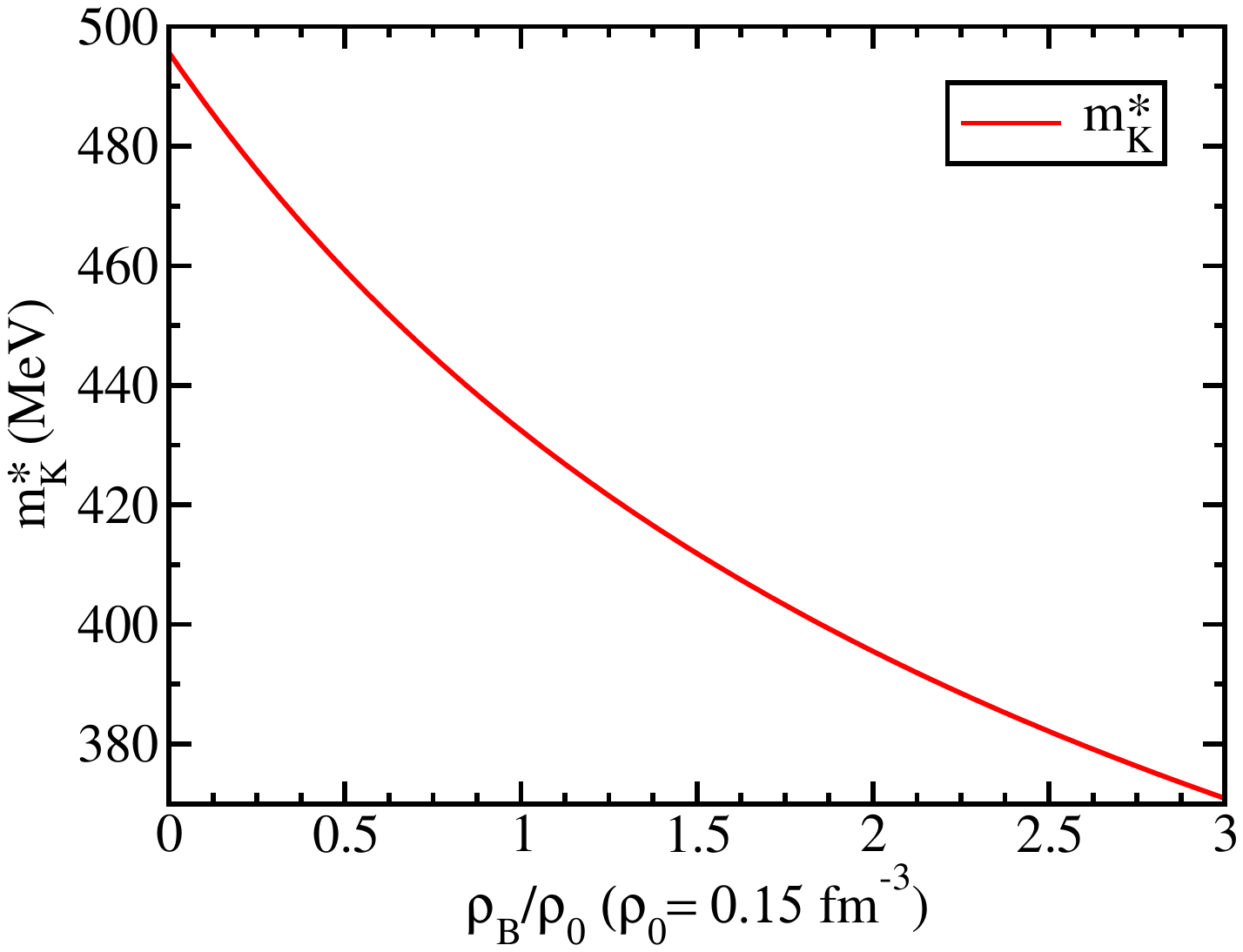}
\caption{\label{fig:mk} In-medium kaon mass $m^*_{K}$.}
\end{figure}

The resulting in-medium kaon (Lorentz-scalar) mass,  
calculated via Eqs.~(\ref{hmass}) and~(\ref{bagRs}), 
is shown in~Fig.~\ref{fig:mk}, with the parameters fixed by the nuclear 
matter saturation properties.  The kaon effective mass at normal nuclear matter density 
$\rho_0 = 0.15$~fm$^{-3}$ decreases by about $13\%$. 
This is a little larger than the 
10\% decrease used in Ref.~\cite{Ko:1992tp}. Note that, the isoscalar-vector 
$\omega$-mean-field potentials arise both for the kaon and antikaon. However, 
they have opposite signs and cancel each other 
(or they can be eliminated by a variable shift)  
in the calculation of the $\phi$ self-energy, 
and therefore we do not show here{\textemdash}see Ref.~\cite{Tsushima:1997df} for details.
%

\section{\label{sec:medse}$\phi$ meson in matter}
The in-medium $\phi$ mass is calculated by solving Eq.~(\ref{eqn:phimassvacuum}) 
by replacing $m_{K}$ by $m_{K}^{*}$ and $m_{\phi}$ by $m_{\phi}^{*}$, and
the width is obtained by using the solutions in Eq.~(\ref{eqn:phidecaywidth}). 
We regularize the associated loop integral with a dipole form factor using a cutoff 
mass parameter $\Lambda_{K}$. In principle, this parameter 
may be determined phenomenologically 
using, for example, a quark model{\textemdash}see Ref.~\cite{Krein:2010vp} for 
more details. 
However, for simplicity we keep it free and vary its value 
over a wide interval, namely
1000-3000~MeV. 
\begin{table}[h]
\begin{center}
\begin{tabular}{l|rrr}
\hline \hline
& $\Lambda_{K}= 1000$ & $\Lambda_{K}= 2000$  & $\Lambda_{K}= 3000$ \\
\hline
$m_{\phi}^{*}$ & 1009.3 & 1000.9 & 994.9 \\
$\Gamma_{\phi}^{*}$ & 37.7 & 34.8 & 32.8 \\
\hline \hline
\end{tabular}
\caption{\label{tab:phippties} $\phi$ mass and width at 
normal nuclear matter density, $\rho_{0}$. All quantities are given in MeV.}
\end{center}
\end{table}

\begin{figure}[htb]
\centering
\includegraphics[scale=0.3]{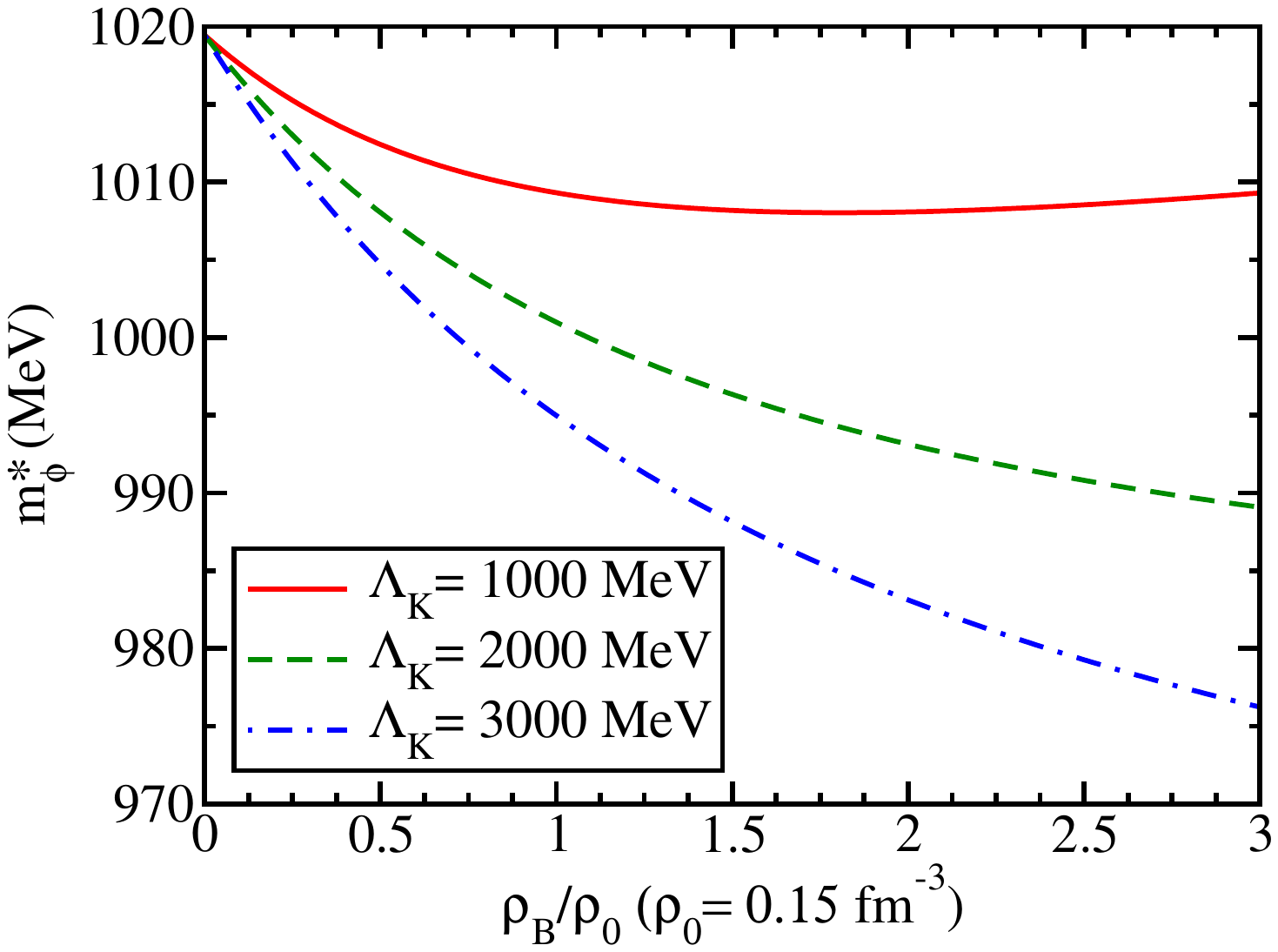}
\includegraphics[scale=0.3]{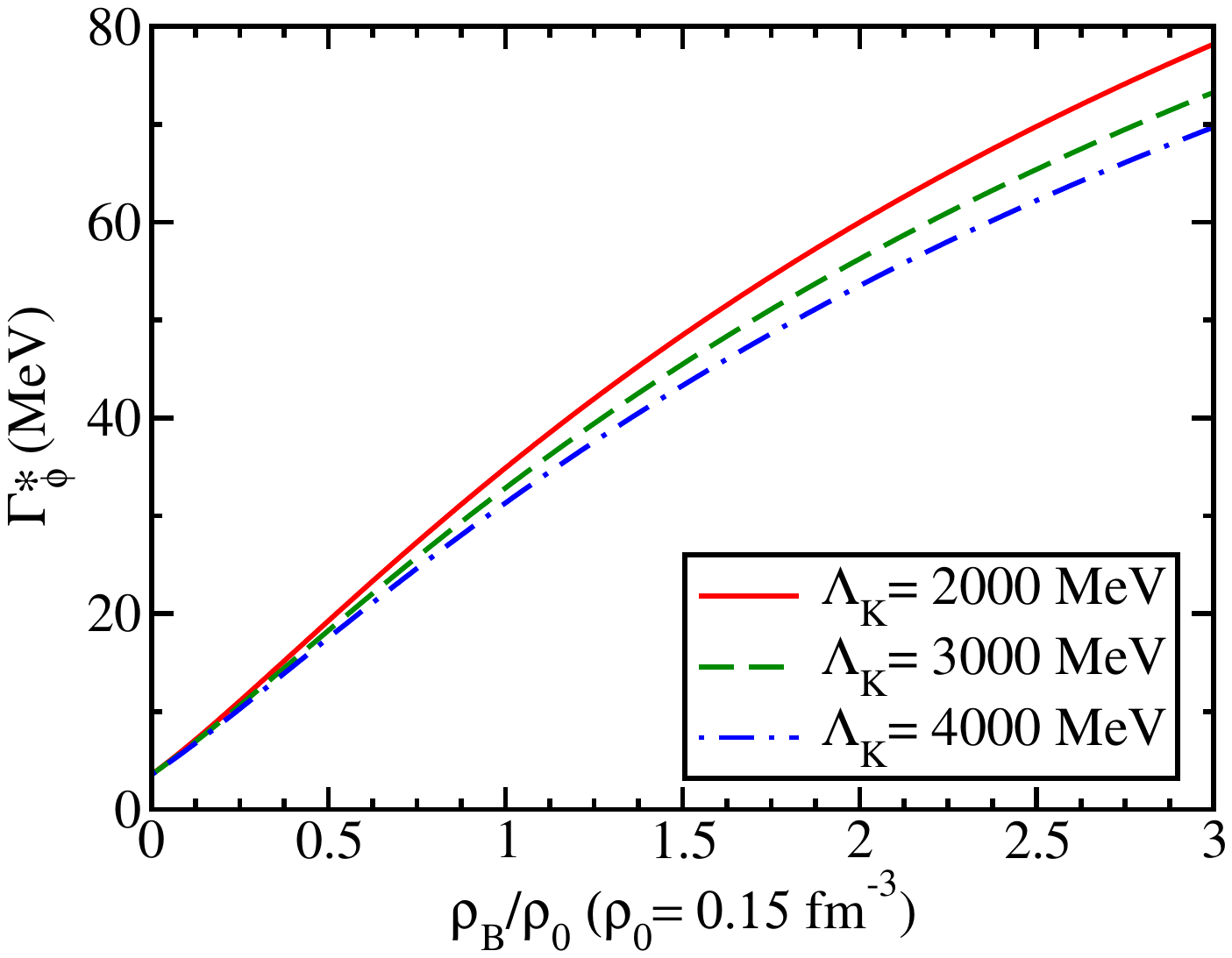}
\caption{\label{fig:mphi} In-medium $\phi$ mass (upper panel) and width 
(lower panel) for three values of the cutoff parameter $\Lambda_{K}$. }
\end{figure}

In~\tab{tab:phippties}, we present the values for $m^*_{\phi}$ and 
$\Gamma_{\phi}^{*}$ at normal nuclear matter density $\rho_{0}$. A negative kaon mass 
shift of 13\% induces only $\approx$ 2\% downward mass shift of the $\phi$. 
On the other hand, $\Gamma_{\phi}^{*}$ is very sensitive 
to the change in the kaon mass; 
at $\rho_B=\rho_0$, the broadening of the $\phi$ becomes 
an order of magnitude larger than 
its vacuum value and it increases rapidly with increasing 
nuclear density, up to a factor of 
$\sim 20$ enhancement for the largest nuclear matter 
density treated, $\rho_{B}=3\rho_{0}$. This can
be seen in Fig.~\ref{fig:mphi}, where we plot $m^*_\phi$ 
and $\Gamma^*_\phi$ as a function
of the ratio $\rho_B/\rho_0$. The effect of the 
in-medium kaon mass change gives a negative 
shift of the $\phi$ meson mass. However, even for 
the largest value of density treated in this 
study, the downward mass shift is only a few percent for 
all values of the cutoff parameter 
$\Lambda_{K}$. For $m_{\phi}^{*}$ at normal nuclear matter density, 
the average downward mass shift is $1.8\%$ with 
a $0.7\%$ standard deviation from the averaged value, 
while $\Gamma_{\phi}^{*}$ broadens in average by a factor of 10 with a 0.7 
standard deviation from the average. 

Next, we present predictions for single-particle energies and half widths for
$\phi$-nucleus bound states for several selected nuclei. We solve the Schr\"odinger
equation for a complex $\phi$-nucleus scalar potential determined by a local-density
approximation using the $\phi$ mass shift and decay width in nuclear matter.
This amounts to using the following for the complex $\phi$-nucleus (A) potential  
\begin{equation} 
V_{\phi A}(r)= \Delta m^*_{\phi}(\rho_{B}(r)) -(\mi/2)\Gamma^*_{\phi}(\rho_{B}(r)),
\end{equation}
where $\Delta m^*_{\phi}(\rho_{B}(r)) \equiv m^*_{\phi}(\rho_{B}(r)) - m_\phi$,   
$r$ is the distance from the center of the nucleus and $\rho_{B}(r)$ is 
the density profile of the given nucleus, which we calculate in the 
QMC model. Table~\ref{tab:phi-nucleus-be} shows the results for the real and imaginary
parts of the single-particle energies $\mathcal{E}=E-(\mi/2)\Gamma$ in $^4{\rm He}$, 
$^{12}{\rm C}$ and $^{208}{\rm Pb}$.
\begin{table}[ht]
\begin{center}
\begin{tabular}{ll|cc|cc}
\hline \hline
  & & E & $\Gamma/2$  & E & $\Gamma/2$ \\
\hline
$^{4}\text{He}$ & 1s & -1.39 & 0 & \multicolumn{2}{c}{n} \\
\hline
$^{12}\text{C}$ & 1s & -7.70 & 0 & -6.47 & 11.00 \\
\hline
$^{208}\text{Pb}$ & 1s & -21.22 & 0 & -21.06 & 16.25 \\
                     & 1p & -17.69 & 0 & -17.35 & 15.76  \\
                     & 1d & -13.34 & 0 & -12.78 & 15.06 \\
                     & 2s & -11.68 & 0 & -10.97 & 14.67  \\
\hline \hline
\end{tabular}
\caption{\label{tab:phi-nucleus-be} $\phi$-nucleus bound state 
single-particle energies $E$ and half widths $\Gamma/2$, calculated 
by solving the Schr\"{o}dinger equation with and without the imaginary part of 
the $\phi$-nucleus potential $V_{\phi A}(r)$. The cutoff value used
is $\Lambda_{K}=3000$ MeV. Quantities are all in MeV. 
``n'' in the entry $^4$He denotes that we find no bound state.
}
\end{center}
\end{table}
We present results with and without the imaginary 
(absorptive) part of the $\phi$-nucleus potential $V_{\phi A}(r)$. One sees that
$\phi$ is not bound to $^{4}{\rm He}$ when the imaginary part of the potential is
included. For larger nuclei, the $\phi$ does bind but 
while the binding is substantial the 
energy levels are quite broad; the half widths being roughly the same size as the 
central values of the real parts. 

\begin{figure}[htb]
\label{fig:xiOn}
\centering
\includegraphics[scale=0.3]{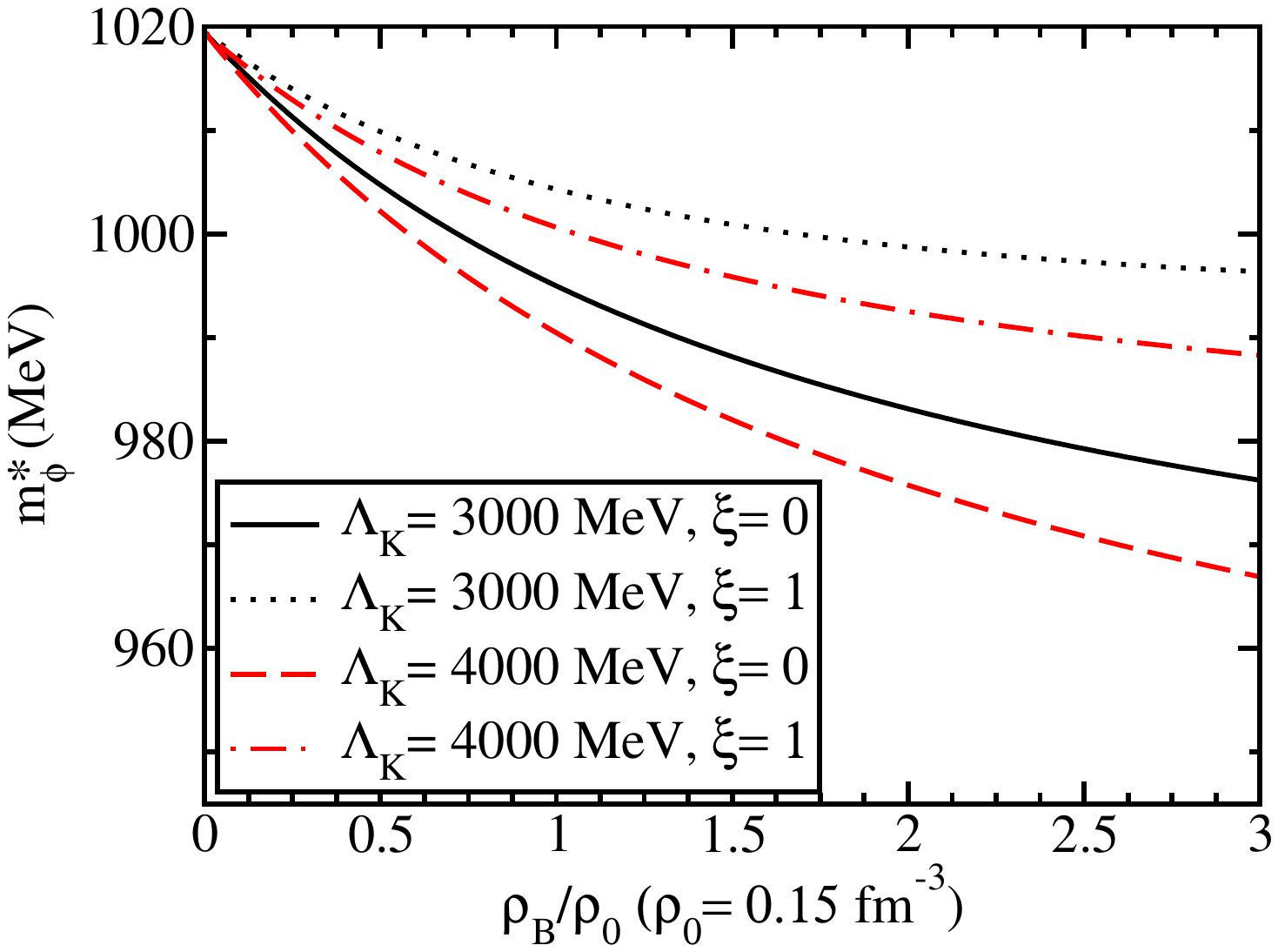}
\includegraphics[scale=0.3]{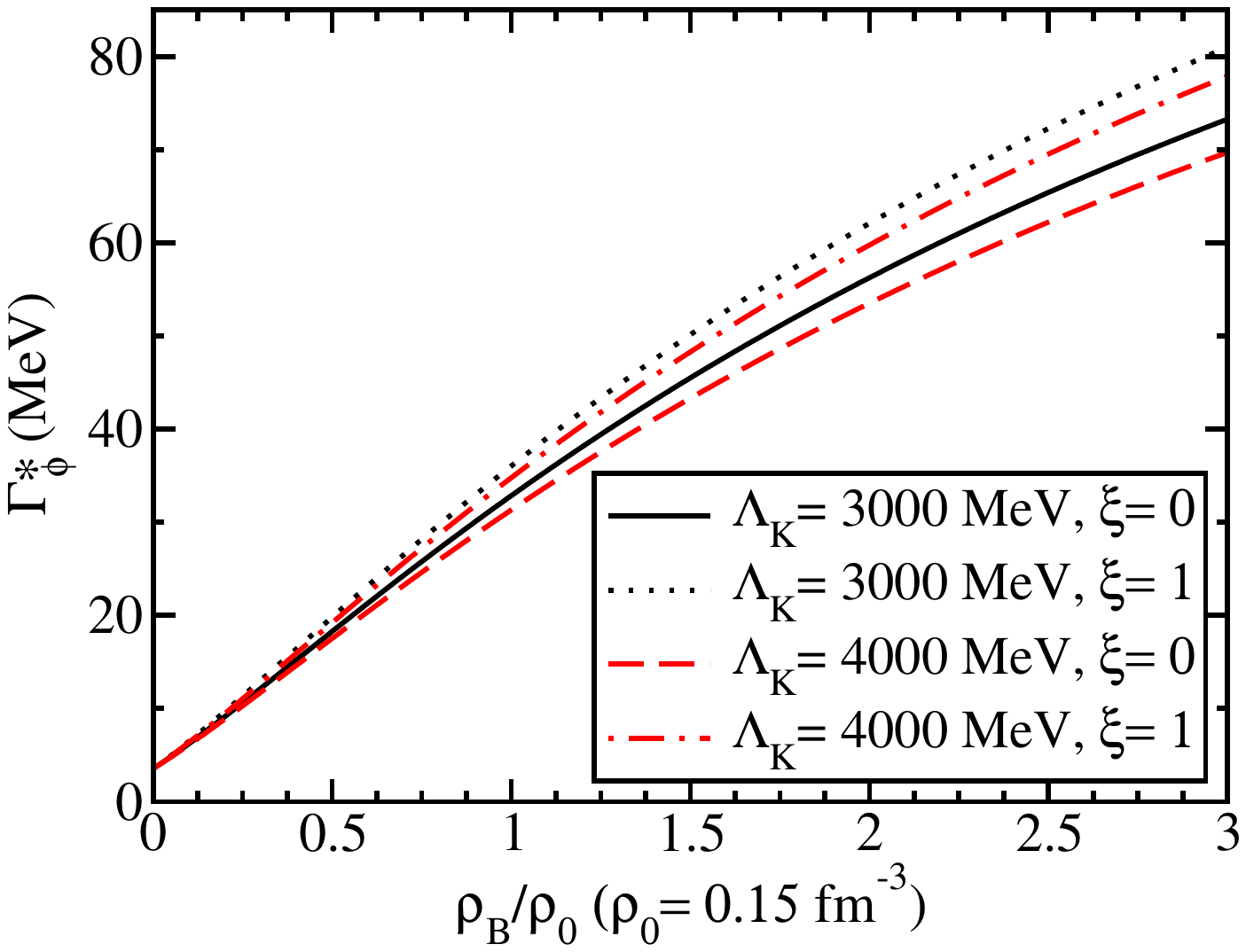}
\caption{Effect of adding ($\xi = 1$) 
the $\phi\phi K \Kbar$ interaction 
of Eq.~(\ref{eqn:phi2kk}) on the in-medium $\phi$ mass (upper panel) 
and width (lower panel) 
for two values of the cutoff parameter $\Lambda_{K}$. }
\end{figure}

To conclude and for completeness, we show the impact of adding the $\phi\phi K \Kbar$ 
interaction of Eq.~(\ref{eqn:phi2kk}) on the 
in-medium $\phi$ mass and width. Figure~\ref{fig:xiOn} 
presents the results. We have used the notation that $\xi = 1 (0)$ means 
that this interaction 
is (not) included in the calculation of the  $\phi$ self-energy. 
One still gets a downward shift 
of the in-medium $\phi$ mass when $\xi = 1$, although the absolute value 
is slightly different from $\xi = 0$. The in-medium width is not
very sensitive to this interaction. 
%

\section{\label{sec:conclusions}Conclusions and perspectives}
We have calculated the $\phi$ meson mass and width in nuclear matter
within an effective Lagrangian approach up to three times  
of normal nuclear matter density.
Essential to our results are the in-medium kaon masses,  
which are calculated in the quark-meson coupling (QMC) 
model, where the scalar and vector meson mean fields couple directly to the 
light $u$ and $d$ quarks (antiquarks) in the $K$ ($\Kbar$) mesons.

At normal nuclear matter density, allowing for a very large variation of the 
cutoff parameter $\Lambda_{K}$, although we have found a sizable 
negative mass shift of 13\% in the kaon mass, this induces only a few percent 
(1.8\% on average) downward shift of the $\phi$ meson mass. 
On the other hand, it induces an order-of-magnitude broadening of the decay width.

Given the nuclear matter results, we have used a local density approximation 
to infer the position dependent 
attractive complex scalar potential, 
$V_{s}(\rho_B(r))= \Delta m_{\phi}^{*}(\rho_B(r))
-(\mi/2)\Gamma_{\phi}^{*}(\rho_B(r))$, in a finite nucleus. This allowed us  
to study the binding
and absorption of a number of $\phi$-nuclear systems,  
given the nuclear density profiles, $\rho_B(r)$, also calculated using 
the QMC model.
While the results found in this study show that one should expect the 
$\phi$ meson to be bound in all but the lightest nuclei, 
the broadening of these energy levels, which is comparable to the amount 
of binding,  may introduce challenges in observing such states  
experimentally.

In the present study, we have focused on the $\phi$ self-energy in medium   
due to the medium modified kaon-antikaon loop.
However, more study of gluonic color forces is needed  
on binding of the $\phi$-meson to a nucleus.

As a possible extension of this work, we note that 
the medium effects on the $\phi$ meson may lead to some enhancement of the 
strangeness content of the bound nucleon, with consequences, for example,
for dark matter detection. 

\vspace{1ex}
\noindent
{\bf Acknowledgments}\\
This work was partially supported by Conselho Nacional de 
Desenvolvimento Cient\'{i}fico e 
Tecnol\'{o}gico - CNPq, Grant Nos. 152348/2016-6 (J.J.C-M.), 400826/2014-3
and 308088/2015-8 (K.T.), 305894/2009-9 (G.K.), and 313800/2014-6 (A.W.T.),  
and Funda{\c c}\~{a}o de Amparo \`{a} Pesquisa do Estado de S\~{a}o Paulo -
FAPESP, Grant Nos. 2015/17234-0 (K.T.) and 2013/01907-0 (G.K.).
This research was also supported by the Australian Research
Council through the ARC Centre of Excellence for Particle
Physics at the Terascale (CE110001104), and through
Grant No. DP151103101 (A.W.T.).


\section*{References}

\begin{thebibliography}{55}
\bibitem{Leupold:2009kz} 
  S.~Leupold, V.~Metag and U.~Mosel,
  Int.\ J.\ Mod.\ Phys.\ E {\bf 19}, 147 (2010)
  [arXiv:0907.2388 [nucl-th]].

\bibitem{Hayano:2008vn} 
  R.~S.~Hayano and T.~Hatsuda,
  Rev.\ Mod.\ Phys.\  {\bf 82}, 2949 (2010)
  [arXiv:0812.1702 [nucl-ex]].

\bibitem{Krein:2016fqh} 
  G.~Krein,
  AIP Conf.\ Proc.\  {\bf 1701}, 020012 (2016).
  
\bibitem{Tsushima:1997df} 
  K.~Tsushima, K.~Saito, A.~W.~Thomas and S.~V.~Wright,
  Phys.\ Lett.\ B {\bf 429}, 239 (1998)
  Erratum: [Phys.\ Lett.\ B {\bf 436}, 453 (1998)]
  [nucl-th/9712044].

\bibitem{Laue:1999yv} 
  F.~Laue {\it et al.} [KaoS Collaboration],
  Phys.\ Rev.\ Lett.\  {\bf 82}, 1640 (1999)
  [nucl-ex/9901005].

\bibitem{SchaffnerBielich:1999cp} 
  J.~Schaffner-Bielich, V.~Koch and M.~Effenberger,
  Nucl.\ Phys.\ A {\bf 669}, 153 (2000)
  [nucl-th/9907095].

\bibitem{Akaishi:2002bg} 
  Y.~Akaishi and T.~Yamazaki,
  Phys.\ Rev.\ C {\bf 65}, 044005 (2002).

\bibitem{Fuchs:2005zg} 
  C.~Fuchs,
  Prog.\ Part.\ Nucl.\ Phys.\  {\bf 56}, 1 (2006)
  [nucl-th/0507017].

  
\bibitem{Titov:1997qz} 
  A.~I.~Titov, Y.~s.~Oh and S.~N.~Yang,
  Phys.\ Rev.\ Lett.\  {\bf 79}, 1634 (1997)
  [nucl-th/9702015].

\bibitem{Titov:1998bw} 
  A.~I.~Titov, Y.~s.~Oh, S.~N.~Yang and T.~Morii,
  Phys.\ Rev.\ C {\bf 58}, 2429 (1998)
  [nucl-th/9804043].

\bibitem{Oh:1999nv} 
  Y.~s.~Oh, A.~I.~Titov, S.~N.~Yang and T.~Morii,
  Phys.\ Lett.\ B {\bf 462}, 23 (1999)
  [nucl-th/9905044].

\bibitem{Oh:2001bq} 
  Y.~s.~Oh and H.~C.~Bhang,
  Phys.\ Rev.\ C {\bf 64}, 055207 (2001)
  [nucl-th/0104068].

\bibitem{Gubler:2014pta} 
  P.~Gubler and K.~Ohtani,
  Phys.\ Rev.\ D {\bf 90}, no. 9, 094002 (2014)
  [arXiv:1404.7701 [hep-ph]].

\bibitem{Bottino:2001dj} 
  A.~Bottino, F.~Donato, N.~Fornengo and S.~Scopel,
  Astropart.\ Phys.\  {\bf 18}, 205 (2002).
  [hep-ph/0111229].

\bibitem{Ellis:2008hf} 
  J.~R.~Ellis, K.~A.~Olive and C.~Savage,
  Phys.\ Rev.\ D {\bf 77}, 065026 (2008).
  [arXiv:0801.3656 [hep-ph]].

\bibitem{Giedt:2009mr} 
  J.~Giedt, A.~W.~Thomas and R.~D.~Young,
  Phys.\ Rev.\ Lett.\  {\bf 103}, 201802 (2009)
  doi:10.1103/PhysRevLett.103.201802
  [arXiv:0907.4177 [hep-ph]].
  
\bibitem{Sibirtsev:2006yk} 
  A.~Sibirtsev, H.~W.~Hammer, U.~G.~Meissner and A.~W.~Thomas,
  Eur.\ Phys.\ J.\ A {\bf 29}, 209 (2006)
  [nucl-th/0606044].

\bibitem{Ishikawa:2004id} 
  T.~Ishikawa {\it et al.},
  Phys.\ Lett.\ B {\bf 608}, 215 (2005)
  [nucl-ex/0411016].

\bibitem{Appelquist:1978rt} 
  T.~Appelquist and W.~Fischler,
  Phys.\ Lett.\  {\bf 77B}, 405 (1978).
  
\bibitem{Brodsky:1989jd} 
  S.~J.~Brodsky, I.~A.~Schmidt and G.~F.~de Teramond,
  Phys.\ Rev.\ Lett.\  {\bf 64}, 1011 (1990).

\bibitem{Luke:1992tm} 
  M.~E.~Luke, A.~V.~Manohar and M.~J.~Savage,
  Phys.\ Lett.\ B {\bf 288}, 355 (1992)
  [hep-ph/9204219].

\bibitem{Sibirtsev:1999jr}
  A.~Sibirtsev, K.~Tsushima, K.~Saito and A.~W.~Thomas,
  Phys.\ Lett.\ B {\bf 484} (2000) 23
  doi:10.1016/S0370-2693(00)00635-3
  [nucl-th/9904015].

\bibitem{Kawama:2014pja} 
  D.~Kawama [J-PARC E16 Collaboration],
  PoS Hadron {\bf 2013}, 178 (2013).

\bibitem{Kawama:2014iwa} 
  D.~Kawama {\it et al.} [J-PARC E16 Collaboration],
  JPS Conf.\ Proc.\  {\bf 1}, 013074 (2014).

\bibitem{Ohnishi:2014xla} 
  H.~Ohnishi {\it et al.},
  Acta Phys.\ Polon.\ B {\bf 45}, 819 (2014).

\bibitem{Aoki:2015qla} 
  K.~Aoki [J-PARC E16 Collaboration],
  arXiv:1502.00703 [nucl-ex].

\bibitem{Morino:2015bqa} 
  Y.~Morino {\it et al.} [J-PARC E16 Collaboration],
  JPS Conf.\ Proc.\  {\bf 8}, 022009 (2015).

\bibitem{Gao:2000az} 
  H.~Gao, T.~S.~H.~Lee and V.~Marinov,
  Phys.\ Rev.\ C {\bf 63}, 022201 (2001)
  [nucl-th/0010042].

\bibitem{Beane:2014sda} 
  S.~R.~Beane, E.~Chang, S.~D.~Cohen, W.~Detmold, H.-W.~Lin, K.~Orginos, A.~Parreño and M.~J.~Savage,
  Phys.\ Rev.\ D {\bf 91}, no. 11, 114503 (2015)
  [arXiv:1410.7069 [hep-lat]].

\bibitem{Brambilla:2015rqa} 
  N.~Brambilla, G.~Krein, J.~Tarrús Castellà and A.~Vairo,
  Phys.\ Rev.\ D {\bf 93}, 054002 (2016)
  [arXiv:1510.05895 [hep-ph]].
  
\bibitem{Gao:2017hya} 
  H.~Gao, H.~Huang, T.~Liu, J.~Ping, F.~Wang and Z.~Zhao,
  arXiv:1701.03210 [hep-ph].


\bibitem{Muto:2005za} 
  R.~Muto {\it et al.} [KEK-PS-E325 Collaboration],
  Phys.\ Rev.\ Lett.\  {\bf 98}, 042501 (2007)
  [nucl-ex/0511019].

\bibitem{Mibe:2007aa} 
  T.~Mibe {\it et al.} [CLAS Collaboration],
  Phys.\ Rev.\ C {\bf 76}, 052202 (2007)
  [nucl-ex/0703013 [NUCL-EX]].

\bibitem{Qian:2009ab} 
  X.~Qian {\it et al.} [CLAS Collaboration],
  Phys.\ Lett.\ B {\bf 680}, 417 (2009)
  [arXiv:0907.2668 [nucl-ex]].

\bibitem{Wood:2010ei} 
  M.~H.~Wood {\it et al.} [CLAS Collaboration],
  Phys.\ Rev.\ Lett.\  {\bf 105}, 112301 (2010)
  [arXiv:1006.3361 [nucl-ex]].

\bibitem{Polyanskiy:2010tj} 
  A.~Polyanskiy {\it et al.},
  Phys.\ Lett.\ B {\bf 695}, 74 (2011)
  [arXiv:1008.0232 [nucl-ex]].

\bibitem{JPARCE16Proposal} 
\url{http://rarfaxp.riken.go.jp/~yokkaich/paper/jparc-proposal-0604.pdf}.

\bibitem{JPARCE29Proposal} 
\url{http://j-parc.jp/researcher/Hadron/en/pac_0907/pdf/Ohnishi.pdf}.

\bibitem{JPARCE29ProposalAdd} 
\url{http://j-parc.jp/researcher/Hadron/en/pac_1007/pdf/KEK_J-PARC-PAC2010-02.pdf}.

\bibitem{JLabphi}
\url{https://www.jlab.org/exp_prog/PACpage/PAC42/PAC42_FINAL_Report.pdf}.

\bibitem{Brown:1991kk}
  G.~E.~Brown and M.~Rho,
  Phys.\ Rev.\ Lett.\  {\bf 66}, 2720 (1991).

\bibitem{Ko:1992tp} 
  C.~M.~Ko, P.~Levai, X.~J.~Qiu and C.~T.~Li,
  Phys.\ Rev.\ C {\bf 45}, 1400 (1992).


\bibitem{Hatsuda:1991ez} 
  T.~Hatsuda and S.~H.~Lee,
  Phys.\ Rev.\ C {\bf 46}, no. 1, R34 (1992).

\bibitem{Hatsuda:1996xt} 
  T.~Hatsuda, H.~Shiomi and H.~Kuwabara,
  Prog.\ Theor.\ Phys.\  {\bf 95}, 1009 (1996)
  [nucl-th/9603043].

\bibitem{Klingl:1997tm} 
  F.~Klingl, T.~Waas and W.~Weise,
  Phys.\ Lett.\ B {\bf 431}, 254 (1998)
  [hep-ph/9709210].

\bibitem{Oset:2000eg}
  E.~Oset and A.~Ramos,
  Nucl.\ Phys.\ A {\bf 679}, 616 (2001)
  [nucl-th/0005046].

\bibitem{Cabrera:2002hc} 
  D.~Cabrera and M.~J.~Vicente Vacas,
  Phys.\ Rev.\ C {\bf 67}, 045203 (2003)
  [nucl-th/0205075].

\bibitem{Gubler:2015yna} 
  P.~Gubler and W.~Weise,
  Phys.\ Lett.\ B {\bf 751}, 396 (2015)
  [arXiv:1507.03769 [hep-ph]].

\bibitem{Cabrera:2016rnc} 
  D.~Cabrera, A.~N.~Hiller Blin and M.~J.~Vicente Vacas,
  Phys.\ Rev.\ C {\bf 95}, no. 1, 015201 (2017)
  [arXiv:1609.03880 [nucl-th]].

\bibitem{Guichon:1989tx} 
  P.~A.~M.~Guichon,
  Nucl.\ Phys.\ A {\bf 497}, 265C (1989).

\bibitem{Guichon:1995ue}
  P.~A.~M.~Guichon, K.~Saito, E.~N.~Rodionov and A.~W.~Thomas,
  Nucl.\ Phys.\ A {\bf 601} (1996) 349
  doi:10.1016/0375-9474(96)00033-4
  [nucl-th/9509034].

\bibitem{Stone:2016qmi}
  J.~R.~Stone, P.~A.~M.~Guichon, P.~G.~Reinhard and A.~W.~Thomas,
  Phys.\ Rev.\ Lett.\  {\bf 116} (2016) no.9,  092501
  doi:10.1103/PhysRevLett.116.092501
  [arXiv:1601.08131 [nucl-th]];\\
  K.~Saito, K.~Tsushima and A.~W.~Thomas,
  Nucl.\ Phys.\ A {\bf 609}, 339 (1996)
  doi:10.1016/S0375-9474(96)00263-1
  [nucl-th/9606020];\\
  K.~Saito, K.~Tsushima and A.~W.~Thomas,
  Phys.\ Rev.\ C {\bf 55}, 2637 (1997)
  doi:10.1103/PhysRevC.55.2637
  [nucl-th/9612001].
  
  
 \bibitem{qmchyp}
  K.~Tsushima, K.~Saito, J.~Haidenbauer and A.~W.~Thomas,
  Nucl.\ Phys.\ A {\bf 630}, 691 (1998)
  doi:10.1016/S0375-9474(98)00806-9
  [nucl-th/9707022];\\
  K.~Tsushima, K.~Saito and A.~W.~Thomas,
  Phys.\ Lett.\ B {\bf 411}, 9 (1997)
  Erratum: [Phys.\ Lett.\ B {\bf 421}, 413 (1998)]
  doi:10.1016/S0370-2693(97)00944-1, 10.1016/S0370-2693(98)00065-3
  [nucl-th/9701047];\\
    K.~Tsushima and F.~C.~Khanna,
  Phys.\ Rev.\ C {\bf 67}, 015211 (2003)
  doi:10.1103/PhysRevC.67.015211
  [nucl-th/0207077];\\
  K.~Tsushima and F.~C.~Khanna,
  J.\ Phys.\ G {\bf 30}, 1765 (2004)
  doi:10.1088/0954-3899/30/12/001
  [nucl-th/0303073];\\
  P.~A.~M.~Guichon, A.~W.~Thomas and K.~Tsushima,
  Nucl.\ Phys.\ A {\bf 814}, 66 (2008)
  doi:10.1016/j.nuclphysa.2008.10.001
  [arXiv:0712.1925 [nucl-th]].
  
\bibitem{Saito:2005rv} 
  K.~Saito, K.~Tsushima and A.~W.~Thomas,
  Prog.\ Part.\ Nucl.\ Phys.\  {\bf 58}, 1 (2007)
  [hep-ph/0506314].


\bibitem{Klingl:1996by} 
  F.~Klingl, N.~Kaiser and W.~Weise,
  Z.\ Phys.\ A {\bf 356}, 193 (1996)
  [hep-ph/9607431].
  
\bibitem{Lin:1999ve} 
  Z.~w.~Lin, C.~M.~Ko and B.~Zhang,
  Phys.\ Rev.\ C {\bf 61}, 024904 (2000).

\bibitem{Lee:1994wx} 
  S.~H.~Lee, C.~Song and H.~Yabu,
  Phys.\ Lett.\ B {\bf 341}, 407 (1995)
  [hep-ph/9408266].

\bibitem{Yamawaki:1986zz} 
  K.~Yamawaki,
  Phys.\ Rev.\ D {\bf 35}, 412 (1987).


\bibitem{Meissner:1986tc} 
  U.~G.~Meissner and I.~Zahed,
  Z.\ Phys.\ A {\bf 327}, 5 (1987).

  
\bibitem{Meissner:1987ge} 
  U.~G.~Meissner,
  Phys.\ Rept.\  {\bf 161}, 213 (1988).

\bibitem{Saito:1987ba} 
  S.~Saito and K.~Yamawaki,
  NAGOYA, JAPAN: UNIV., PHYS. DEPT. (1987) 225p

\bibitem{Krein:2010vp} 
  G.~Krein, A.~W.~Thomas and K.~Tsushima,
  Phys.\ Lett.\ B {\bf 697}, 136 (2011)
  [arXiv:1007.2220 [nucl-th]].

  \bibitem{PDG:2015} 
  The Review of Particle Physics (2015),
 K.A. Olive et al. (Particle Data Group), 
 Chin. Phys. C, 38, 090001 (2014) and 2015 update,
 \url{http://pdg.lbl.gov/}









\end{thebibliography}
\end{document}